\documentclass[twocolumn,showkeys,showpacs,preprintnumbers,prd,superscriptaddress,nofootinbib]{revtex4}
\usepackage{graphicx}
\usepackage{epsf}
\usepackage{bm}
\usepackage{amsmath}
\usepackage{amsfonts}
\usepackage{amssymb}
\usepackage{epstopdf}
\usepackage{natbib}
\usepackage{color}
\usepackage{verbatim}
\usepackage{caption}
\usepackage{multirow}
\usepackage{bm}
\usepackage{listings}
\usepackage{soul}
\usepackage{caption}
\usepackage{subcaption}
\usepackage{ragged2e}
\usepackage{float}
\usepackage{hyperref}

\begin{document}

\title{An investigation of a varying G through Strong Lensing and SNe Ia observations}

\author{R. F. L. Holanda}
\email{holandarfl@gmail.com}
\affiliation{Universidade Federal do Rio Grande do Norte, Departamento de F\'{i}sica Te\'{o}rica e Experimental, 59300-000, Natal - RN, Brazil.}

\affiliation{Departamento de F\'{\i}sica, Universidade Federal de Campina Grande, 58429-900, Campina Grande - PB, Brasil}

\author{Marcelo Ferreira}
\email{fsm.fisica@gmail.com}
\affiliation{Universidade Federal do Rio Grande do Norte, Departamento de F\'{i}sica Te\'{o}rica e Experimental, 59300-000, Natal - RN, Brazil.}

\author{Javier E. Gonzalez} \email{javiergonzalezs@academico.ufs.br}
\affiliation{Departamento de F\'{i}sica, Universidade Federal de Sergipe, São Cristóvão, SE 49100-000, Brazil}



\begin{abstract}
\noindent 
In this paper, we analyze the potential variation of the gravitational constant $G$ using data from strong gravitational lensing systems and Type Ia supernovae. Testing $G(z)$ parameterizations where $G(z) = G_0(1 + G_1z)$ and $G(z) = G_0(1 + z)^{G_1}$, we also account for the influence of $G$ on the luminosity of SNe Ia through the Chandrasekhar mass-luminosity relation. Only the flat universe hypothesis is considered. Constraints from 158 lensing systems and the Pantheon+ sample show no significant evidence of $G$ variation. However, although the results are compatible with no variation, the errors are not yet sufficiently restrictive to rule out any variation of $G$ with high statistical confidence. This study highlights the viability of using combined astrophysical data to probe variations in fundamental constants, suggesting that future surveys could refine these constraints.

\end{abstract}



\maketitle

\section{Introduction}
The $\Lambda$CDM model, as the current cosmological standard, has shown remarkable consistency with most of the observational probes available so far. However, it remains theoretically imperfect, with several unresolved challenges and observational tensions \cite{Efstathiou:2024dvn,Lin:2019htv,Gonzalez:2021ojp}. The most significant among them is the Hubble tension, which refers to the notable discrepancy between the Hubble constant ($H_0$) derived from local distance ladder measurements and that inferred from the cosmic microwave background. This tension poses a critical challenge to the standard $\Lambda$CDM paradigm. Recent studies have placed this issue at the forefront of modern cosmological research, incorporating a wide array of new data and interpretations. In parallel, the broader pursuit of understanding the universe has led to increasing interest in the potential variation of fundamental constants. These constants—such as the Newtonian gravitational constant ($G$), the fine-structure constant ($\alpha$), and the speed of light in vacuum ($c$)—are foundational pillars of physics. Although traditionally assumed to be invariant across time and space, theoretical advancements and observational anomalies have reignited discussions on their possible variation (see detailed reviews in \cite{Uzan:2010pm,2017RPPh...80l6902M,2023Univ....9...94H,2024arXiv241007281U}).\footnote{Actually,  since Dirac's pioneering work in 1937 \cite{Dirac:1937ti}, where he proposed that large dimensionless numbers could be derived from combinations of these constants, the idea of their variation has gained attention in various contexts.}

For the fine structure constant ($\alpha$), significant advancements have been made in studying its potential spatial and temporal variation, particularly through high-redshift absorption lines in quasar spectra \cite{Dzuba:1998au,Dzuba:1999zz,Webb:1998cq}. Numerous observational data sets and methods have since refined these studies \cite{King:2012id,Galli:2012bf,Leefer:2013waa,vandeBruck:2015rma,Kotus:2016xxb,Colaco:2019fvl,Goncalves:2019xtc,Liu:2021mfk,Bora:2020sws,2024EPJC...84.1120F,2024PhRvD.109j3521G}. In particular, Ref. \cite{2024EPJC...84.1120F} investigated possible cosmological variations in the fine structure constant ($\alpha$) using galaxy cluster gas mass fractions and type Ia supernovae (SNe Ia) data. The results showed that $\alpha$ variation is detected at $3\sigma$ c.l. for $z \lesssim 0.5$ when adopting Planck $H_0$, while local $H_0$ measurements restore consistency with a constant $\alpha$. The variation of the speed of light ($c$) has also been intensely debated due to its implications for both cosmology and the foundations of physics \cite{Moffat93,Albrecht:1998ir,Magueijo:2003gj,Pedram:2007mj,Chen:2008uv}. Models exploring this possibility, often referred to as Varying Speed of Light (VSL) theories, suggest that modifications in the Maxwell-Einstein framework could allow for light propagation at speeds exceeding the metric-defined limit. However, such theories introduce challenges to causality and quantum mechanics \cite{Teyssandier:2003qh,Adams:2006sv,Lee:2020zts,2018PDU....22..127C,Xu:2016zxi,Liu:2018qrg,Zhu:2021pml,Xu:2016zsa}. Recent investigations of quantum gravity effects, such as those analyzing the travel time of cosmic photons, indicate that $c$ may depend on the photon energy \cite{Li:2022szn}. The impact of a variable speed of light on early-universe physics and cosmology has been a topic of extensive discussion for decades \cite{Dicke:1957zz,Avelino:1999is,Magueijo:2000zt,Moffat:2014poa,2024JCAP...11..062S,2024arXiv241205262N,2022JCAP...07..029R}.\footnote{Simultaneous variation of multiple constants has also been explored. For example, the interdependence of variations in $G$, $c$, and $\Lambda$ (cosmological constant) has been investigated in the context of the Covarying Physical Couplings (CPC) framework \cite{Franzmann:2017nsc,Costa:2017abc,Cuzinatto:2022dta,Cuzinatto:2022mfe,Gupta:2021tma,Gupta:2022amf}.}

With respect to the gravitational constant, several recent papers have explored the implications for  current tensions in the standard model by considering possible temporal variations of $G$, both in phenomenological analyses and within some theoretical framework \cite{2020EPJC...80..570W,2022JCAP...06..004B,PhysRevD.97.083505,2020EPJC...80..148A,2021Univ....7..366A,2024arXiv240803875R,2024JCAP...06..056R,2021PhRvD.104b1303M,2024arXiv240715553L}. For instance, the Ref. \cite{2024JCAP...06..056R}  proposed a gravitational transition model that modifies the gravitational constant \( G \) at distances below 50 Mpc, affecting the Cepheid Period-Luminosity relation and the absolute magnitude of SNe Ia. With \( \Delta G/G \approx 0.04 \), the model naturally resolved the Hubble tension by providing a \( H_0 \) value consistent with Planck measurements. The Ref.  \cite{2021PhRvD.104b1303M}  showed that the Hubble tension can be addressed by a rapid transition in the effective gravitational constant, \( \mu_G \equiv G_{\text{eff}} / G_N \), at \( z_t \simeq 0.01 \). This transition decreases the luminosity of SNe Ia by 0.2~mag, resolving the \( H_0 \) and growth tensions without altering the Planck/\(\Lambda\)CDM expansion history. On the other hand, the Ref. \cite{2024arXiv240715553L},  using data from the latest Planck PR4 release and DESI BAO measurements,  achieved no significant variation for $G$. Their result being robust across various assumptions about the cosmological model, including non-standard dark energy fluids and non-flat models. However, the situation differed if, for instance, both the fine structure constant and $G$ were allowed to vary simultaneously. 

In this paper, we show that measurements of the Einstein radius in strong gravitational lensing systems combined with SNe Ia observations provide a pathway to test a possible variation in the gravitational constant without relying on any specific cosmological model, focusing solely on the assumption of a flat universe. We employ the following hypotheses to test a possible variation of
$G$: $G(z)=G_0(1+G_1z)$ and $G(z) = G_0(1 + z)^{G_1}$. We consider a sample of 158 confirmed sources of Strong Gravitational Lensing
 obtained from SLOAN
Lens ACS \cite{2008ApJ...682..964B, 2009ApJ...705.1099A}, BOSS Emission-line Lens Survey \cite{2012ApJ...744...41B}, Strong Legacy Survey SL2S \cite{2011ApJ...727...96R, 2013ApJ...777...98S, 2013ApJ...777...97S, 2015ApJ...800...94S} and SLACS \cite{shu2017sloan}. The SNe Ia luminosity distance measurements are obtained by using
the Pantheon+ sample. However, the exact connection between Chandrasekhar mass  and Type Ia supernova luminosity remains a topic of debate (see, for instance, Ref. \cite{2024JCAP...06..056R}). Although it was traditionally believed that luminosity increases with mass \cite{woosley1986physics, 1999astro.ph..7222A, gaztanaga2001bounds}, recent studies propose an inverse correlation \cite{wright2018type, 2019PhRvD.100j4035S}. Therefore, we have performed our method considering the two hypotheses. It is very important to mention that the Friedmann equations are modified if $G$ varies over time. However, our methodology does not require the use of such equations.


This paper is organized as follows. The methodology adopted in this work is presented in section \ref{Methodology}. In section \ref{samples} we describe the cosmological data used in our analysis. The analysis and main results are presented in section \ref{results}. Finally, we conclude in section \ref{final}.






\section{Methodology}
\label{Methodology}

Strong gravitational lensing (SGL), one of the remarkable predictions of General Relativity, has emerged as a powerful astrophysical tool for probing gravitational and cosmological theories, measuring key cosmological parameters, and exploring fundamental physics \cite{2004mmu..symp..117K, 2017JCAP...07..010R, gott1989setting, 2019MNRAS.483.1104Q, 2018ApJ...866...31R, 2017JCAP...09..039H, 2018ApJ...867...50C, 2023ApJ...948...47H, 2024arXiv241117888H, 2024PhRvD.109h4074L}. These phenomena occur when the source (\(s\)), lens (\(l\)), and observer are nearly aligned in a straight line. 


In systems where a single galaxy acts as the lens, the Einstein radius depends on three primary factors: the angular diameter distances to both the lens and the source, and the mass distribution within the lens galaxy \cite{Leaf_2018}. 
Based on the assumption of the Singular Isothermal Sphere (SIS) model,  the Einstein radius is given by \cite{schneider1992gravitational, 2015ApJ}

\begin{equation} \label{eq1}
    \theta_E = 4 \pi \frac{\sigma^2_{SIS}}{c^2} \Delta,
\end{equation}
with

\begin{equation}
    \Delta \equiv \frac{D_{A,ls}}{D_{A,s}},
\end{equation}
where $D_{A,ls}$ and $D_{A,s}$ are the angular diameter distances between lens and source and between source and observer, respectively, and $\sigma_{SIS}$ is the velocity dispersion,
which depends on the gravitational constant as \cite{schneider1992gravitational} $\sigma^2_{SIS} = M G/r$, where $M$ is the lens mass and $r$ is a characteristic scale of the system. So, it is clear that $ \sigma^2_{SIS} \propto G.$ Therefore, allowing G to vary as $G(z) = G_0 \phi(z)$, we correct equation \eqref{eq1} as

\begin{equation} \label{thetae_1}
    \theta_E = \phi(z_l) \left( 4 \pi \frac{\sigma^2_{SIS, 0}}{c^2} \Delta \right),
\end{equation}
where $\phi(z)$ is responsible for a possible $G$ variation with $z$, with $z_l$ representing the redshift of the lens and $\sigma^2_{SIS, 0}$ is the velocity dispersion evaluated considering $G_0$. { In purely gravitational systems, observables like the velocity dispersion $\sigma$ depend on the product $GM$, making it challenging to disentangle $G$ from $M$ without independent, non-gravitational mass estimates. However, our approach is to investigate the relative evolution of the velocity dispersion $\sigma(z)$ as a potential indicator of a varying gravitational coupling $G(z)$, rather than attempting to extract absolute values of $G$. Since $\sigma^2 \sim GM/r$, both $M$ and $r$ are expected to evolve with redshift due to structure formation processes. {   In principle, in a statistically large and morphologically diverse sample, the redshift evolution of structural properties such as \( M(z) \) and \( r(z) \) may average out, reducing their impact on the observable \( \sigma(z) \). In such cases, residual trends in velocity dispersion might reflect additional effects — potentially including a varying gravitational coupling \( G(z) \). This effect could be even more pronounced if the lensing population shares similar dynamical properties or lies within a narrow mass range.  This work explores, for the first time, whether residual trends in $\sigma(z)$—as currently observed in the literature—might offer statistical sensitivity to a possible variation in $G(z)$. Our analysis opens a novel observational avenue to test the constancy of the gravitational coupling over cosmological timescales.}



We can assume a flat universe, where the comoving distance between the lens and source is $r_{ls} = r_s - r_l$, and employing the relations $r_l = D_{A,l} (1+z_l)$ and $r_s = D_{A,s} (1+z_s)$, to obtain the following \cite{2021EPJC...81..822C, 2022JCAP...08..062C}:

\begin{equation}
    \Delta = 1 - \frac{(1 + z_l)}{(1+z_s)} \frac{D_{A,l}}{D_{A,s}},
\end{equation}
where $z_s$ is the redshift of the source.
 Additionally, we can consider the CDDR:
 
\begin{equation}
    D_A(z) = \frac{D_L(z)}{(1 + z)^2},
\end{equation}
 to write it in terms of the luminosity distance, $D_L$, as follows:

\begin{equation}
\label{D_DL}
    \Delta = 1 - \frac{(1 + z_s)}{(1+z_l)} \frac{D_{L,l}}{D_{L,s}}.
\end{equation}
Hence, we combine Eqs. \eqref{thetae_1} and \eqref{D_DL} into the expression: 

\begin{equation} \label{eq_phi1}
    D_0 = \phi(z_l) \left[ 1 - \frac{(1 + z_s)}{(1+z_l)} \frac{D_{L,l}}{D_{L,s}} \right],
\end{equation}
{where $D_0  = \theta_E c^2/(4 \pi \sigma_{SIS}^2)$ is the observational quantity.} As mentioned earlier, our methodology does not rely on the Friedmann equations, which are modified in a scenario where $G$ may vary with time.

{ It is worth to comment that the CDDR holds in all cosmological models based on Riemannian geometry and does not rely on Einstein's field equations or on the specific properties of the universe’s matter-energy content. Actually, three fundamental conditions must be satisfied: spacetime must be described by a metric theory of gravity; photons must travel along null geodesics; and the number of photons must be conserved \cite{Etherington33,Ellis07}. In the last decade, different methods have been proposed to test the validity of the CDDR due to the improvement in the quantity and the quality of astronomical data \cite{2012IJMPD..2150008H,Rasanen:2015kca,Renzi:2021xii,Teixeira:2025czm,Qi:2024acx,Keil:2025ysb,Rana:2015feb,Gahlaut:2025lhv,Mukherjee:2021kcu,More:2016fca}. Its validity has been verified, at least, within $2\sigma$ c.l. Very recently, in the Ref.  \cite{2025arXiv250410464T}, the authors explored whether deviations from the CDDR could alleviate the Hubble tension and influence constraints on dark energy parameters derived from cosmological data. In a related study, it is assessed how potential CDDR violations might affect the determination of the Hubble constant, $H_0$ \cite{2012IJMPD..2150008H}. More recently, the authors in Ref. \cite{2025arXiv250416868A} employed a consistency test based on the DDR to identify possible systematic effects in DESI DR2 and Pantheon+ datasets, aiming to ensure the reliability of dark energy measurements.  Nevertheless, several observational tests based on strong gravitational lensing samples have been performed, and no significant deviation from the CDDR has been detected. Therefore, for the dataset used in the present work, the validity of the relation is readily assumed \cite{2022EPJC...82..115H,2025ApJ...979....2Q,2017JCAP...07..010R,2017JCAP...09..039H,2016JCAP...02..054H}. }

\subsection{Chandrasekhar mass and
the luminosity of Type Ia supernovae}

Now, in order to obtain the luminosity distances to the lens and to the source using  SNe Ia, we need to examine how the evolution of $G$ influences the luminosity of SNe Ia. Type Ia supernovae are used as standard candles in cosmology due to the well-established relationship between their intrinsic luminosities and light curves. However, the evolution of fundamental physical parameters, such as the gravitational constant $G$, can modify this relationship, thus affecting the measurements of the cosmological distance. In this context, we analyze how the evolution of $G$ affects the luminosity of SNe Ia, particularly through its dependence on Chandrasekhar mass.

The relationship between the Chandrasekhar mass and the luminosity of SNe Ia remains debated. Although it was traditionally thought that the luminosity increases with $M_{ch}$, recent studies suggest an inverse relationship. According to some authors, the peak luminosity of SNe Ia is proportional to the Chandrasekhar mass of the exploding white dwarf, such that $L_{\text{SNe}} \propto M_{ch}$ \cite{woosley1986physics, 1999astro.ph..7222A, gaztanaga2001bounds}. Given that $M_{ch} \propto G^{-3/2}$, this implies an inverse relationship between $G$ and $M_{ch}$, leading to a decrease in the luminosity of SNe Ia as $G$ increases. 
This inverse relationship can be understood as follows \cite{2024JCAP...06..056R}: if $G$ is lower than its usual value, the gravitational force per unit mass decreases. This reduction allows the electron degeneracy pressure to resist the gravitational pull of a larger mass just before collapse. Consequently, the Chandrasekhar mass increases for a lower $G$. 

Given this dependence, {considering the relation between luminosity and luminosity distance, i.e, $D_L = \sqrt{L / 4\pi F}$
, where $F$ is the observed flux}, the measured luminosity distance from SNe Ia should be corrected using the following expression:

\begin{equation}
    D_L^{\text{SNe}} = D_{L,0}^{\text{SNe}} \phi(z)^{-3/4},
\end{equation}

\noindent where $D_{L,0}^{\text{SNe}}$ represents the luminosity distance measured under the assumption of a constant $G$, and $\phi(z)$ encodes the evolution of $G$ with redshift.

In theoretical frameworks involving non-standard gravity, it has been argued that the standardized SNe Ia peak luminosity decreases with an increase in the Chandrasekhar mass rather than increasing, as might be intuitively expected \cite{wright2018type}. This counterintuitive result arises from factors such as the dynamics of the explosion and the interaction of radiation with the surrounding ejecta. A semi-analytic model of the SNe Ia light curves, developed by Wright and Li \cite{wright2018type} (see also Ref. \cite{2019PhRvD.100j4035S}), suggests a scaling relation of the form $L_{\text{SNe}} \propto M_{ch}^{-0.97}$. Combining this result with $M_{ch} \propto G^{-3/2}$, we find that $L_{\text{SNe}} \propto G^{1.46}$. 

Therefore, we also  adopt this possibility in our analysis, assuming that the evolution of $G$ directly impacts both the Chandrasekhar mass and the observed luminosities of SNe Ia such as 

\begin{equation}
    D_L^{\text{SNe}} = D_{L,0}^{\text{SNe}} \phi(z)^{0.73}.
\end{equation} 

Taking into account these changes in SNe luminosity, we should correct the equation \eqref{eq_phi1} as 

\begin{equation} \label{D_phi}
    D_0 = \phi(z_l) \left[ 1 - \frac{(1 + z_s)}{(1+z_l)} \frac{D_{L,l,0} \phi(z_l)^{n}}{D_{L,s,0}\phi(z_s)^{n}} \right],
\end{equation}
where $n = -3/4$ for the first case and $n = 0.73$ for the second one.

We consider the following models for $\phi(z)$:

\begin{equation}
    \phi(z)=\begin{cases}1+G_1z & \text{(model }P_1) \\
    (1+z)^{G_1} & (\text{model } P_2)
    \end{cases}
    \label{phi1}
\end{equation}  
where $G_1$ is a free parameter. 
Naturally, if $G_1=0$ is obtained from the analyzes, a constant $G$ is recovered. It is important to emphasize that the above
expressions are phenomenological. {   The models considered in Eq. \eqref{phi1} correspond to the parameterizations:
\( G(z) = G_0 (1 + G_1 z) \) and \( G(z) = G_0 (1 + z)^{G_1} \). The linear parametrization \( G(z) = G_0(1 + G_1 z) \) is adopted as a first-order Taylor expansion around \( z = 0 \). Although strictly valid at low redshift, we extend its use up to \( z \sim 1 \) as a phenomenological ansatz to test for possible deviations from a constant gravitational coupling within the current data. The power-law form remains smooth and well-behaved up to \( z \sim 1 \) and 
is widely used in cosmological analyses to capture mild redshift evolution. Similar parametrizations have also been considered in different contexts, such as tests of the cosmic distance-duality relation \cite{2010ApJ...722L.233H,2021ApJ...909..118Z,2020PhRvD.102f3513D,2019ApJ...885...70L,2018MNRAS.480.3117L,2017IJMPD..2650097F,2024arXiv240803875R,2024JCAP...06..056R}. Additionally, alternative functional forms for \( G(z) \) have been explored in various other theoretical scenarios \cite{2006PhRvD..73j3511N,2009PhRvD..79j4006N,2011A&A...530A..68T,2020JCAP...12..019H,2017arXiv170310538N,2022JCAP...06..004B}. Our goal is not to propose a fundamental model for \( G(z) \), but to explore whether current data exhibit any statistically significant deviation from a constant gravitational coupling.

}





\section{Samples} \label{samples}

In the following, we describe the datasets that will be applied in the methodology presented in the previous section.

\begin{itemize}

\item{\bf SNe Ia Sample:}

We utilize a set of luminosity distances from the Pantheon+ sample \cite{Scolnic:2021amr}. This sample consists of 1701 light curves of 1550 spectroscopically confirmed SNe Ia within a redshift range of $0.001 \leq z \leq 2.26$. The luminosity distance for each supernova is determined through the relation:

\begin{equation}
    D_L = 10^{(m_b - M_b - 25)/5} \ \text{Mpc},
\end{equation}
 where $m_b$ and $M_b$ are the apparent and absolute magnitudes, respectively. In order to perform an independent cosmological model test, we considered the local estimate, $M_b = - 19.253 \pm 0.027$ from \cite{Scolnic:2021amr, riess2022comprehensive}.

 In addition, we use the Gaussian Process (GP) method trained on the SNe Ia sample to reconstruct $D_L(z)$ (see fig. \ref{fig:sne}) at the redshifts $z_l$ and $z_s$ of the SGL systems. This approach allows us to extract  $D_L$ values specifically at the lens and source redshifts.  GP reconstruction is carried out by selecting a prior mean function and a covariance kernel, as detailed in \cite{Seikel:2012uu}. The covariance kernel is defined by a set of hyperparameters that govern the properties of the reconstructed function and quantifies the correlation between the values of the dependent variable in the reconstruction. In this work, we adopt a zero prior mean function to minimize bias in the reconstruction. For the covariance kernel, we use the standard Gaussian kernel, expressed as:

 \begin{equation} \label{gaussian_kernel}
     k(z,z') = \sigma^2 \exp \left(  - \frac{(z - z')^2}{2l^2}
     \right),
 \end{equation}
where $\sigma$ and $l$ are hyperparameters that control the variability of the reconstructed function and its characteristic smoothing scale, respectively. The values of these hyperparameters are determined by maximizing the logarithm of the marginal likelihood:

\begin{equation*}
\ln {\mathcal{L}}=-\frac{1}{2}([\bm { D_{L}}])^T [\bm K(\bm z,\bm z)+\bm C]^{-1}[\bm { D_{L}}]
\end{equation*}

\begin{equation}
\label{log_likelihood}
  -\frac{1}{2}\ln |\bm K(\bm z,\bm z)+\bm C|-\frac{N}{2}\ln 2\pi.
\end{equation}

Here, \( \bm{z} \) represents the vector of redshift measurements from the Pantheon+ dataset, and \( \bm{K}(\bm{z}, \bm{z}) \) denotes the covariance matrix of the data described as a Gaussian Process. The elements of \( \bm{K}(\bm{z}, \bm{z}) \) are computed using Eq.~(\ref{gaussian_kernel}), such that \( [\bm{K}(\bm{z}, \bm{z})]_{i,j} = k(z_i, z_j) \). In addition, \( \bm{C} \) is the covariance matrix of the data and \( N \) is the total number of data points in the sample. We use the GaPP code\footnote{\url{https://github.com/carlosandrepaes/GaPP}} to perform the GP reconstruction and an own code to determine the correlations of the reconstruction as describe in Ref. \cite{Gonzalez:2024qjs}. With this information, we perform a MonteCarlo samplig to calculate the ratios $D_{L,l}/D_{L,s}$ of each SGL system and their covariances, $[C_{D_{L,l}/D_{L,s} }]_{ij}= \text{cov}([D_{L,l}/D_{L,s}]_i,[D_{L,l}/D_{L,s}]_j)$ .

\begin{figure}[htbp]
\includegraphics[width=0.45\textwidth]{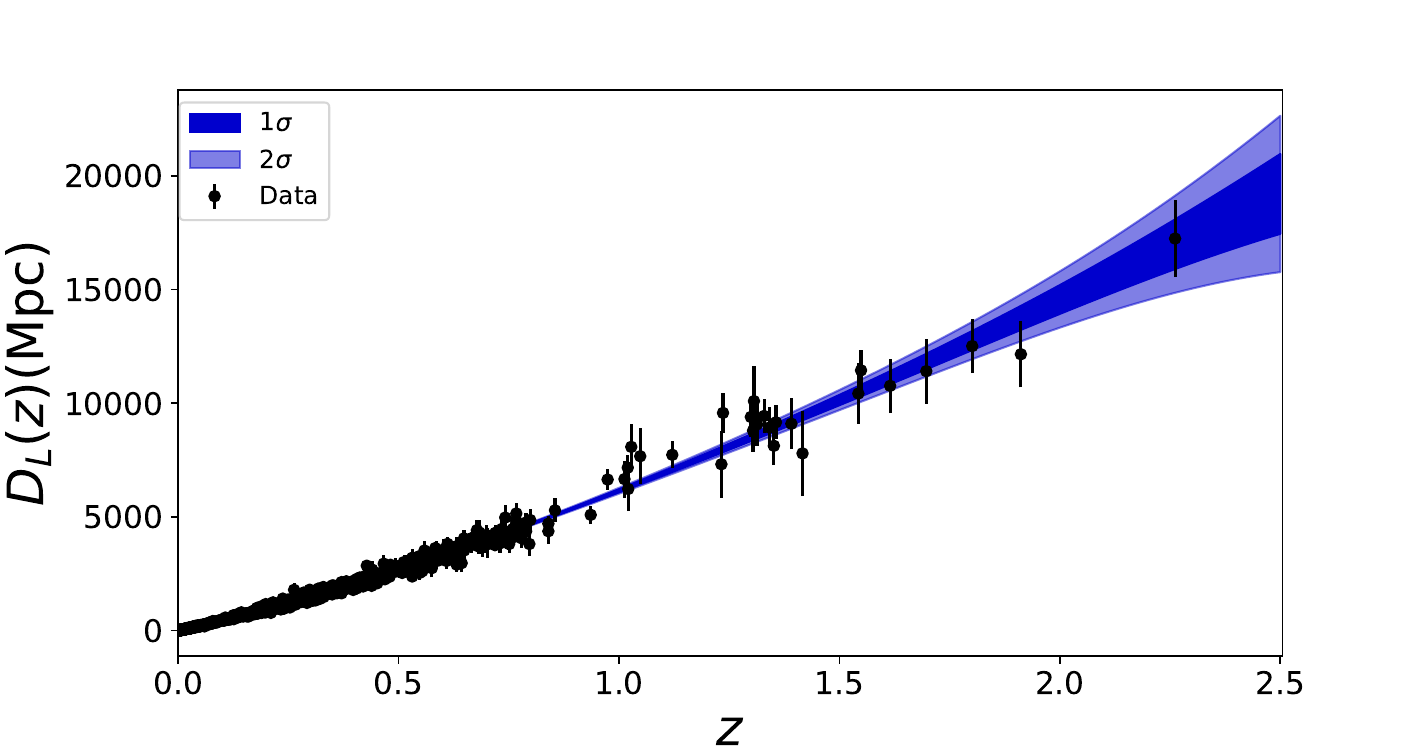}
\caption{Reconstruction of $D_L(z)$ from the Pantheon+ Sample.}
\label{fig:sne}
\end{figure}

\begin{figure*}[htbp]
\centering
\includegraphics[scale=0.4]{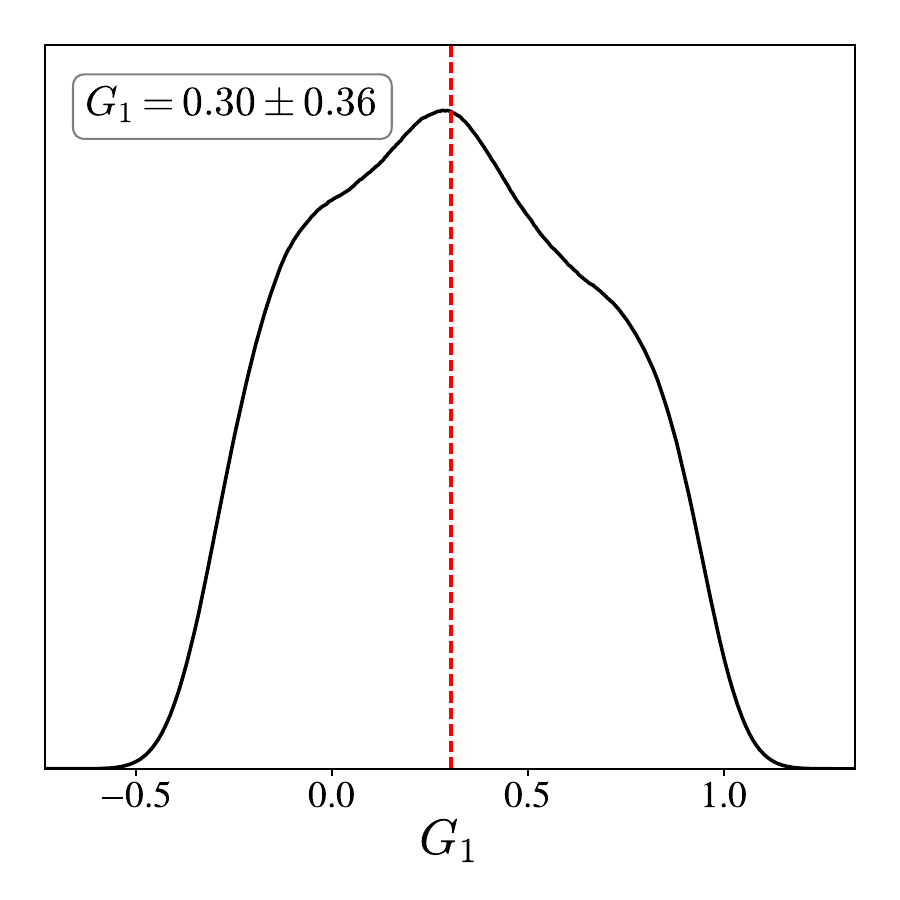}
\includegraphics[scale=0.4]{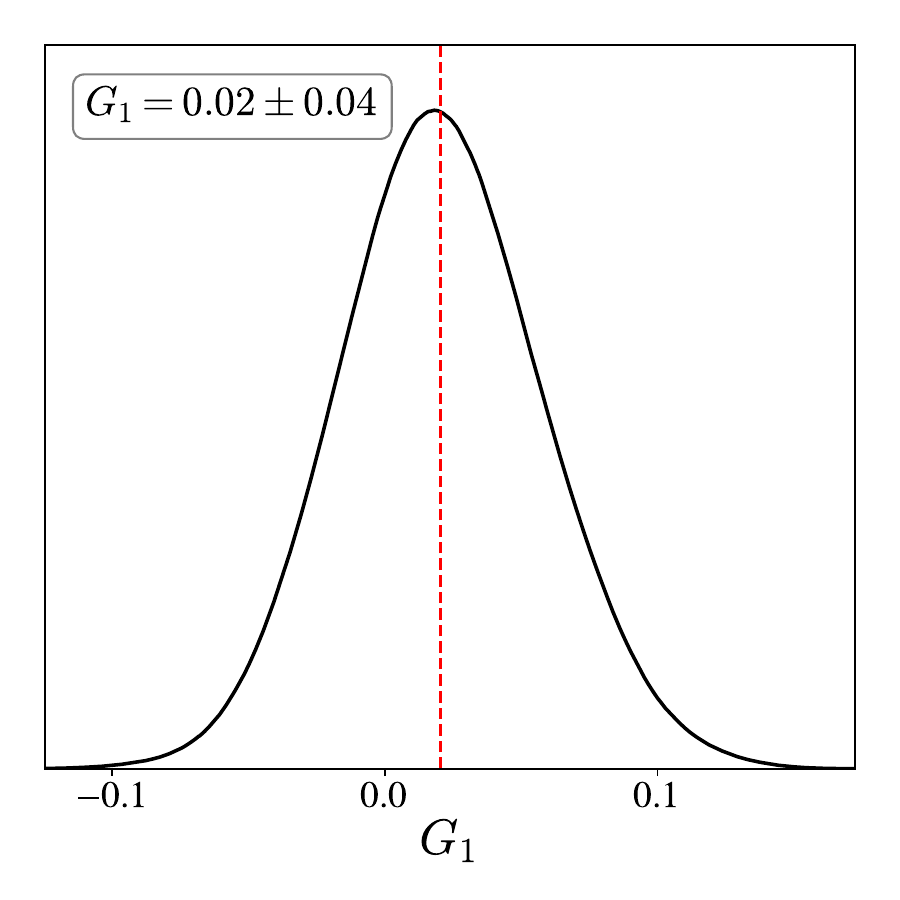}
\caption{By considering: $G(z)=(1+G_1z)$ \textbf{(model $P_1$)}. Left: Posteriori probability distribution of $G_1$ by considering $n = -3/4$. Right: Posteriori probability distribution of $G_1$ by considering $n = 0.73$.}
\label{fig:3/4}
\end{figure*}

\item{\bf Strong Gravitational Lensing Systems:}

In this work, we consider a sample of 158 confirmed sources of Strong Gravitational Lensing from the Ref. \cite{Leaf_2018}, which are compilations of \cite{2015ApJ} and \cite{shu2017sloan}. These systems were obtained from SLOAN Lens ACS \cite{2008ApJ...682..964B, 2009ApJ...705.1099A}, BOSS Emission-line Lens Survey (BELLS) \cite{2012ApJ...744...41B}, Strong Legacy Survey SL2S \cite{2011ApJ...727...96R, 2013ApJ...777...98S, 2013ApJ...777...97S, 2015ApJ...800...94S} and SLACS \cite{shu2017sloan}. The redshifts of these systems are all determined spectroscopically. {In fact, all the observational quantities used in this work are presented in Table I of the Ref. \cite{Leaf_2018}. It is worth mentioning that, instead of directly adopting $\sigma_{\rm SIS}$, we follow the procedure outlined by the Ref. \cite{2015ApJ}, in which the observed stellar velocity dispersion $\sigma_{\rm ap}$, measured within a specific aperture, is rescaled to represent the dispersion within a circular aperture corresponding to half the effective radius of the lens galaxy. This correction is applied through the relation $\sigma_0 = \sigma_{\rm ap} (\theta_{\rm eff} / 2\theta_{\rm ap})^{-0.04}$, where $\theta_{\rm eff}$ is the angular half-light radius and $\theta_{\rm ap}$ denotes the aperture size used in the spectroscopic measurement. The use of $\sigma_0$ makes our observable $D_{0}$ more homogeneous for the sample of lenses located at different redshifts \cite{2015ApJ}. As previously mentioned, all the observational quantities are presented in Table I of the Ref. \cite{Leaf_2018}.  Then, we must use the following.} 

\begin{equation} \label{D0}
    D_0 = \frac{\theta_E c^2}{4 \pi \sigma_0^2}.
\end{equation}
The angular radius of the strong gravitational lens, $\theta_E$, in this sample ranges from \(0.36''\) to \(2.55''\), while the velocity dispersion for this sample spans from \(125\, \mathrm{km\,s^{-1}}\) to \(391\, \mathrm{km\,s^{-1}}\). The uncertainty related to \eqref{D0} is given by

\begin{equation}
    \sigma_{D_0} = D_0 \sqrt{ \left( \frac{\sigma_{\theta_E}}{\theta_E}\right)^2 + \left(2 \frac{\sigma_{\sigma_0}}{\sigma_0}\right)^2 + \sigma_x^2},  
\end{equation}
 where, following \cite{Grillo_2007}, $\sigma_{\theta_E} = 0.05 \theta_E$ (i.e, $5 \%$ for all systems), and $\sigma_x$ represents the intrinsic scatter associated with the SIS model for the lens, approximately $12 \%$, as reported by \cite{Leaf_2018}.   

Additionally, we exclude some systems in which non-physical values for $D_0$ are obtained ($D_0 > 1$), and those with source redshifts outside the supernova redshift range. After these exclusions, we retained 112 systems spanning the redshift ranges $0.0625 \leq z_l \leq 0.722$ and $0.196 \leq z_s \leq 2.15$. {With steps similar to those presented here, Ref.~\cite{Leaf_2018} found results fully consistent with Planck when such models (\(\Lambda\)CDM and \(w\)CDM) were analyzed (see Table II of that paper), despite the still large uncertainties. 
}

\end{itemize}

\section{Analysis and Main Results}
\label{results}

We employed the Markov Chain Monte Carlo (MCMC) method to calculate the posterior probability distribution function (PDF) of the free parameter, using the \texttt{ emcee} sampler \cite{2013PASP125306F}. The constraints were derived by maximizing the likelihood function, defined as:

\begin{equation}
\label{likelihood}
    \mathcal{L} = \frac{1}{\sqrt[n]{2 \pi} |C_\text{tot}|^{1/2}} \exp \left( - \frac{1}{2} \chi^2 \right),
\end{equation}
where the chi-squared is given by  
\begin{equation}
    \chi^2 = (D_0 - \xi)C_\text{tot}^{-1}(D_0 - \xi)^{T}
\end{equation}

and  

\begin{equation} \label{xi}
    \xi \equiv \phi(z_l) \left[ 1 - \frac{(1 + z_s)}{(1 + z_l)} \frac{D_{L,l} \phi(z_l)^{n}}{D_{L,s}\phi(z_s)^{n}} \right].
\end{equation}
The total uncertainty covariance matrix,  \(C_{\text{tot}}\), is calculated as $C_{\text{tot}} = C_{D_0} + C_{\xi}$, where $C_{D_0}$ is the diagonal matrix of square errors of $D_0$ data and the terms of the $\xi$-covariance matrix   are calculate as:

\begin{eqnarray*}
   [C_{\xi} ]_{ij}= \phi(z_{l,i}) \frac{(1 + z_{s,i})}{(1 + z_{l,i})} \frac{\phi(z_{l,i})^{n}}{\phi(z_{s,i})^{n}} \times 
   \end{eqnarray*}
   \begin{eqnarray}
   \label{C_xi}
   \times [C_{D_{L,l}/D_{L,s} }]_{ij}  \frac{\phi(z_{l,j})^{n}}{\phi(z_{s,j})^{n}}\frac{(1 + z_{s,j})}{(1 + z_{l,j})}\phi(z_{l,j}).
\end{eqnarray}

\noindent As mentioned above, we assume $\phi(z) = 1 + G_1 z$ ($P_1$) and $\phi(z_s) = (1 +  z)^ {G_1}$ ($P_2$), where \(G_1\) is a free parameter. The PDF is proportional to the product of the likelihood and the prior:  

\begin{equation}
    P(G_1\mid \text{data}) \propto \mathcal{L}(\text{data} \mid G_1) \times P_0(G_1),
\end{equation}
where we adopt a flat prior within the range \(-0.5 \leq G_1 \leq 1.5\). As shown in Eq \eqref{C_xi}, the $C_\xi$ depends on the $G_1$ parameter, consequently, $C_{\text{tot}}$ also depends on it. For this reason, the normalization factor, $1/|C_\text{tot}|^{1/2}$, in the calculation of the likelihood in Eq. \eqref{likelihood} can not be neglected.

Our main results are summarized as follows:

\begin{itemize}
    \item For \textbf{(model $P_1$)}: $n = -3/4$ in Eq.~\eqref{D_phi}, corresponding to $L_{\text{SNe}} \propto M_{ch}$, we find $G_1 = 0.30 \pm 0.36$ at the $1\sigma$ confidence level. This result is illustrated in Fig.~\ref{fig:3/4}, with a reduced chi-squared value of $\chi^2_{\text{red}} \approx 0.93$. For $n = 0.73$ in Eq.~\eqref{D_phi}, corresponding to $L_{\text{SNe}} \propto M_{ch}^{-0.97}$, we find $G_1 = 0.02 \pm 0.04$ at the $1\sigma$ confidence level. This result is shown in Fig.~\ref{fig:3/4}, also with $\chi^2_{\text{red}} \approx 0.93$.
    \item For \textbf{(model $P_2$)}:  $n=-3/4$ and  $n=0.73$, the following values are obtained (fig. \ref{fig:097}):  $G_1= 0.36 \pm 0.35$ and $G_1=0.03 \pm 0.05$, respectively. Both with $\chi^2_{red} \approx 0.93$. These results are fully consistent with the previous ones. 
\end{itemize}

\noindent { As one may see,  we obtain good values of the reduced chi-squared in our fits. This factor supports the robustness of the filtering step and the good quality of the resulting dataset.} Despite relatively weak limits, these results are consistent with standard scenarios at the confidence level $68\%$, where $G$ is assumed to remain constant. The substantial uncertainties in both cases highlight the challenges in tightly constraining $G_1$ within this framework. However, the reduced chi-squared value of $ \chi^2_{\text{red}} \approx 0.93 $ for both models, indicates a reasonable agreement between the models and the observational data. 

{ A more general power-law scaling \( D_L \propto G^n \) is an interesting extension. We performed a preliminary analysis, but current strong lensing data are not sufficient for reliable constraints. This remains a promising direction once better data become available. {  \noindent
Our results are fully consistent with a vanishing time variation of the gravitational coupling today, i.e., 
\(
|\dot{G}/G|_0 = 0
\), within \(1\sigma\) uncertainties. Although cosmological data still lack the constraining power to detect small deviations, we find no significant evidence for a time-varying \(G\). Independent bounds from local measurements—such as lunar laser ranging, pulsar timing, and helioseismology—are considerably tighter, typically at the level of 
\(
|\dot{G}/G|_0 \lesssim 10^{-13}~\text{yr}^{-1}
\)~\cite{uzan2003fundamental,2014LRR....17....4W,2019MNRAS.482.3249Z,2024arXiv241007281U}. These local constraints, while much stronger, complement cosmological observations by testing the constancy of fundamental couplings across different regimes.
}

\begin{figure*}[htbp]
\centering
\includegraphics[scale=0.4]{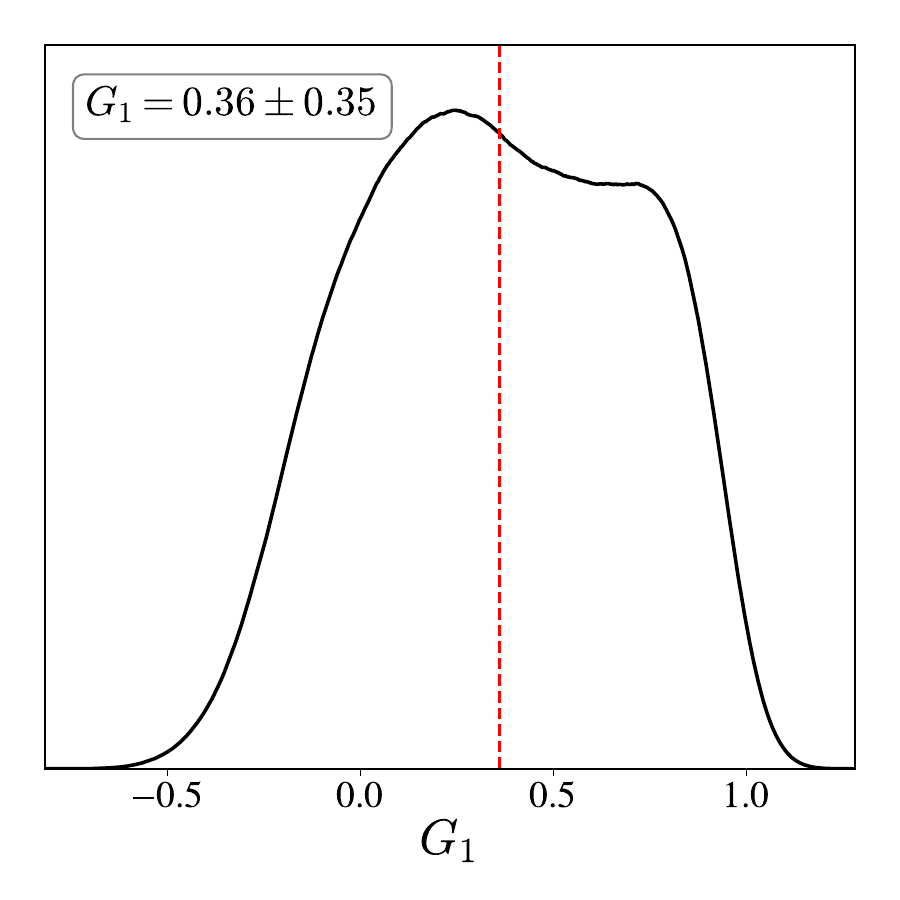}
\includegraphics[scale=0.4]{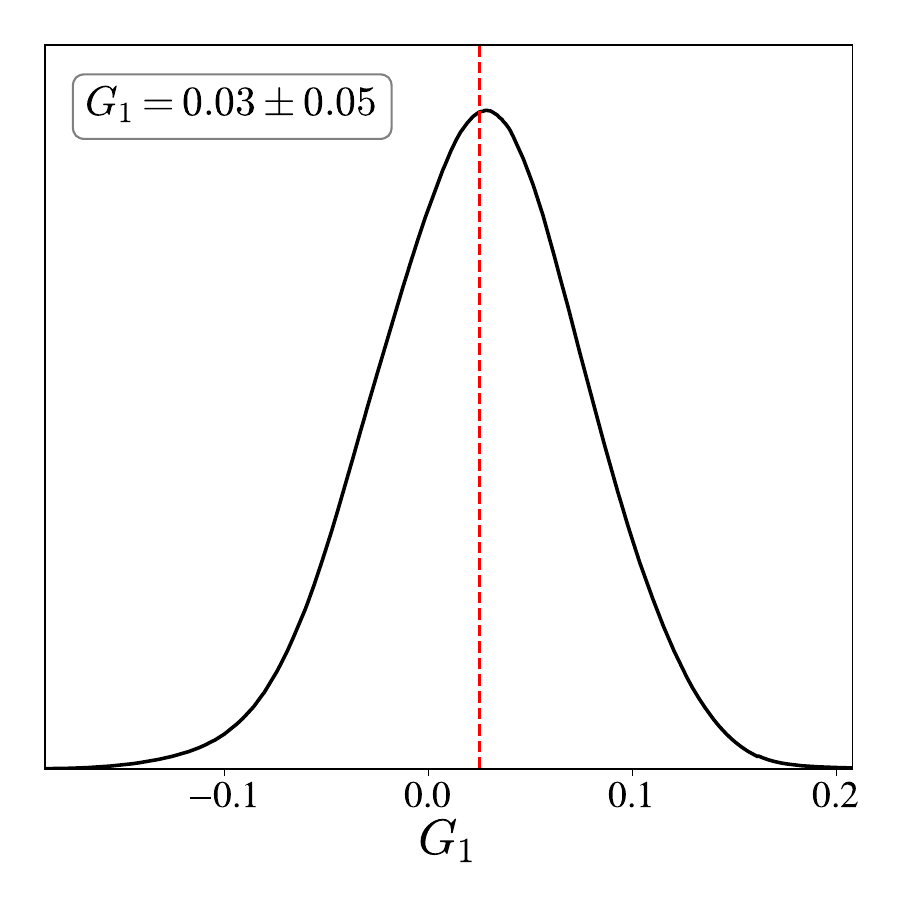}
\caption{By considering: $G(z)=(1+z)^{G_1}$ \textbf{(model $P_2$)}. Left: Posteriori probability distribution of $G_1$ by considering $n = -3/4$. Right: Posteriori probability distribution of $G_1$ by considering $n = 0.73$.}
\label{fig:097}
\end{figure*}

\section{Conclusions}
\label{final}

Some works have explored the possibility of resolving tensions in the standard cosmological model, particularly the Hubble tension, by introducing variations in fundamental constants like, for instance, the gravitational constant $G$. Then, in this paper, we investigated a possible temporal variation of the gravitational constant $G$, using strong gravitational lensing and Type Ia supernova observations. By analyzing Einstein radius measurements and luminosity distances, we tested two hypotheses about the redshift dependence of $G$. In addition, although it was traditionally believed that luminosity increases with mass, recent studies propose an inverse correlation. In this work, we performed our method considering the two hypotheses. Moreover, the methodology presented here did not require the use of the Friedmann equations. Our results show no significant evidence for a variation in $G$. The parameter $G_1$, which describes the variation as $G(z) = G_0 (1 + G_1 z)$, remains consistent with a constant $G$ at the confidence level $68\%$ for both hypotheses. Specifically, we found $G_1 = 0.30 \pm 0.36$ for the hypothesis of decreasing luminosity with increasing $G$, and $G_1 = 0.02 \pm 0.04$ for the alternative scenario. Furthermore, we also examined the parameterization $G(z) = G_0(1+z)^{G_1}$, obtaining results that align with the previous analysis. We derived, respectively, $G_1 = 0.03 \pm 0.05$, and $G_1 = 0.36 \pm 0.35$.  Our results are compatible with no variation in $G$; however, given the still large error bars, they do not exclude recent attempts to resolve the $H_0$ tension by introducing small variations in $G$ in the local universe.

{{ Although our analysis is restricted to a spatially flat universe—consistent with current observational evidence—extending it to include spatial curvature is, in principle, a meaningful direction. However, given that the uncertainties on the parameter of interest are already substantial in the flat case, introducing curvature as an additional free parameter would significantly weaken the constraints. For this reason, we defer such a generalization to future work, when higher-precision data from upcoming surveys will enable a more robust treatment. Nonetheless, the present methodology already highlights the potential of combining lensing and supernova data to probe variations in fundamental constants. The next generation of telescopes—such as Euclid, the Nancy Grace Roman Space Telescope, and the Vera C. Rubin Observatory—is expected to dramatically increase the number and quality of available systems. Additionally, the James Webb Space Telescope will play a crucial role in advancing these efforts (see \cite{treu2022strong} for further discussion). As both statistical and systematic uncertainties in SGL observations are reduced, the framework proposed here may offer increasingly stringent and minimally model-dependent constraints on a possible variation of $G$. A better understanding of the redshift evolution of the lens properties, particularly the typical mass $M(z)$ and scale radius $r(z)$, may further enhance the sensitivity of this approach to probe gravitational coupling evolution.}
}


\begin{acknowledgments}
\noindent {  We thank the referee for the insightful comments and suggestions, which have significantly improved the quality of the manuscript.} RFLH thanks the financial support from the Conselho
Nacional de Desenvolvimento Cientıfico e Tecnologico (CNPq) under the project No. 308550/2023-47.

\end{acknowledgments}

\bibliography{PRD}

\end{document}